\documentclass[]{spie}  

\usepackage{amsmath,amsfonts,amssymb}
\usepackage{graphicx}
\usepackage[colorlinks=true, allcolors=blue]{hyperref}
\usepackage{float}
\usepackage{subcaption}
\usepackage{setspace}

\renewcommand{\vec}[1]{\mathbf{#1}}
\newcommand{\vecs}[1]{\boldsymbol{#1}}

\title{Markov-Chain Monte Carlo Approximation of the Ideal Observer using Generative Adversarial Networks}

\author[a]{Weimin Zhou}
\author[b]{Mark A. Anastasio}
\affil[a]{Department of Electrical and Systems Engineering,
\break Washington University in St$.\ $Louis, St$.\ $Louis, MO 63130, USA}
\affil[b]{Department of Bioengineering, 
\break University of Illinois at Urbana-Champaign, Urbana, IL 61801, USA}

\authorinfo{Further author information: (Send correspondence to Mark A. Anastasio)\\Mark A. Anastasio: E-mail: maa@illinois.edu, Telephone: 1 217 333 1867}

\begin{document} 
\maketitle

\begin{abstract}
The Ideal Observer (IO) performance has been advocated
when optimizing medical imaging systems for signal detection tasks.
However, analytical computation of the IO test statistic is generally intractable. To approximate the IO test statistic, sampling-based methods that employ Markov-Chain Monte Carlo (MCMC) techniques have been developed. However, current applications of MCMC techniques have been limited to several object models such as a lumpy object model and a binary texture model, and it remains unclear how MCMC methods can be implemented with other more sophisticated object models.
Deep learning methods that employ generative adversarial networks (GANs) hold great promise to learn stochastic object models (SOMs) from image data. In this study, we described a method to approximate the IO by applying MCMC techniques to SOMs learned by use of GANs. The proposed method can be employed with arbitrary object models that can be learned by use of GANs, thereby the domain of applicability of MCMC techniques for approximating the IO performance is extended. In this study, both signal-known-exactly (SKE) and signal-known-statistically (SKS) binary signal detection tasks are considered. The IO performance computed by the proposed method is compared to that computed by the conventional MCMC method.
The advantages of the proposed method are discussed.
\end{abstract}


\section{INTRODUCTION}
\label{sec:intro}  
It has been widely accepted that task-based measures of image quality (IQ) should be employed for assessing and optimizing medical imaging systems\cite{barrett2013foundations}.
Task-based measures of IQ quantify the ability of an observer to perform specific tasks\cite{kupinski2001ideal,barrett2013foundations,zhou2018learning, zhou2019learning,zhou2019learningHO,zhou2019approximating}. When optimizing imaging systems for maximizing the amount of task-specific information in the measured images, the performance of the Ideal Observer (IO) 
can be employed as a figure-of-merit (FOM)\cite{barrett2013foundations, kupinski2001ideal,  zhou2018learning, zhou2019learning, zhou2019approximating}. However, for binary signal detection tasks, the IO test statistic is computed by calculating the likelihood ratio that is generally analytically intractable. To approximate the IO test statistic, sampling-based methods using Markov-Chain Monte Carlo (MCMC) techniques \cite{kupinski2003ideal, park2003ideal} have been developed. 
However, current applications of MCMC techniques are limited to relatively simple stochastic object models (SOMs) such as the lumpy background model\cite{kupinski2001ideal} and binary texture model\cite{abbey2008ideal}, and it remains unclear how MCMC methods can be implemented with other more sophisticated object models.

Deep-learning methods employing generative adversarial networks (GANs) \cite{goodfellow2014generative, zhou2019learningSOM} 
hold great promise to learn SOMs that describe 
the variability in the class of objects to-be-imaged.
GANs comprise a generator and a discriminator. By playing a two-player minimax game between the generator and the discriminator, the distribution learned by the generator can approximate
the distribution corresponding to the training data \cite{goodfellow2014generative}. One subsequently can generate new images by sampling latent vectors that constitute the input to the generator. A latent vector is typically a low-dimensional random vector that follows
a simple distribution such as the normal distribution or uniform distribution. 

In this study, inspired by the MCMC algorithm developed by Kupinski \emph{et al.}\cite{kupinski2003ideal}, we propose a novel methodology for approximating the IO by applying MCMC techniques to SOMs learned by use of GANs. Specifically, a GAN is trained on a set of object images to establish a SOM, and MCMC techniques are subsequently applied to the GAN-represented SOM to compute the likelihood ratio. 
As a proof-of-concept, a lumpy background model and an idealized parallel-hole collimator system were considered. Both signal-known-exactly (SKE) and signal-known-statistically (SKS) binary signal detection
tasks were considered. Receiver operating characteristic (ROC) curves and the area under the ROC curve (AUC) values corresponding to the proposed MCMC-GAN algorithm are compared to those corresponding to the conventional MCMC method.  
The potential advantages of the proposed method are discussed.

\section{Background}
Consider a binary signal detection task that requires an observer to classify an image $\vec{g}$ as satisfying a signal-absent hypothesis ($H_0$) or a signal-present hypothesis ($H_1$). 
The imaging processes can be represented as:
\begin{subequations}
\label{eq:imgH_s}
\begin{align}
H_{0}:&\ \mathbf{g} = \mathbf{b} + \mathbf{n}, \\
H_{1}:&\  \mathbf{g} = \mathbf{b} + \mathbf{s} + \mathbf{n},
\end{align}
\end{subequations}
where $\mathbf{b} \in \mathbb{R}^{M}$ denotes an image of background, $\mathbf{s} \in \mathbb{R}^{M}$ denotes the signal to be detected, and $\mathbf{n}\in \mathbb{R}^{M}$ denotes the random measurement noise.

\subsection{Ideal Observer and Markov-Chain Monte Carlo techniques}
The Ideal Observer (IO) sets an upper performance limit among all observers. The IO test statistic 
can be computed as any monotonic transformation of the likelihood ratio:
\begin{equation}
\Lambda (\vec{g})  = \frac{p(\vec{g}|H_1)}{p(\vec{g}|H_0)}.
\end{equation}
However, computation of $\Lambda (\vec{g})$ generally is intractable analytically. 

Kupinski \emph{et al.} proposed a method to numerically approximate the IO test statistic by employing MCMC techniques \cite{kupinski2003ideal}.
For a signal-known-exactly (SKE) binary signal detection task, the likelihood ratio can be written as\cite{kupinski2003ideal}:
\begin{equation} \label{eq:mcmc}
\Lambda (\vec{g})  = \frac{\int d\vec{b}\ p_{b}(\vec{b}) p(\vec{g}|\vec{b}, H_1)}{\int d\vec{b}\ p_{b}(\vec{b}) p(\vec{g}|\vec{b}, H_0)} \equiv \int d\vec{b}\ \Lambda_{\text{BKE}} (\vec{g}|\vec{b}) p(\vec{b}|\vec{g}, H_0),
\end{equation}
where $\Lambda_{\text{BKE}} (\vec{g}|\vec{b}) = \frac{p(\vec{g}|\vec{b},H_1)}{p(\vec{g}|\vec{b},H_0)}$ and $p(\vec{b}|\vec{g}, H_0) = \frac{p(\vec{g}|\vec{b}, H_0) p_b(\vec{b})}{\int d\vec{b'} p(\vec{g}|\vec{b'}, H_0) p_b(\vec{b'})}$. 
The BKE likelihood ratio $\Lambda_{\text{BKE}} (\vec{g}|\vec{b})$ sometimes has an analytical form that is dependent on the type of measurement noise\cite{kupinski2001ideal}.
In cases where the background can be described by a stochastic object model (SOM) with a set of stochastic parameters $\vecs{\theta}$, i.e., $\vec{b} \equiv \vec{b}(\vecs{\theta})$,  the likelihood ratio described in Eq. \ref{eq:mcmc} can be written as\cite{kupinski2003ideal}:
$\Lambda (\vec{g})  = \int d\vecs{\theta}\ \Lambda_{\text{BKE}} (\vec{g}|\vec{b}(\vecs{\theta})) p(\vecs{\theta}|\vec{g}, H_0).$
Subsequently, the likelihood ratio can be approximated as:
\begin{equation}
\hat{\Lambda}(\vec{g}) = \frac{1}{J}\sum_{j=1}^{J} \Lambda_{\text{BKE}} (\vec{g}|\vec{b}(\vecs{\theta}^{j})).
\end{equation}
Here, each $\vecs{\theta}^{j}$ is sampled from the posterior distribution $p(\vecs{\theta}|\vec{g}, H_0)$. To sample $\vecs{\theta}^{j}$ from the distribution $p(\vecs{\theta}|\vec{g}, H_0)$, 
a Markov chain with the stationary density $p(\vecs{\theta}|\vec{g}, H_0)$ can be generated. To do this, an initial vector $\vecs{\theta}^{0}$ is chosen and a 
proposal density function $q(\vecs{\theta} | \vecs{\theta}^j)$ is specified. Given $\vecs{\theta}^j$, the candidate vector $\hat{\vecs{\theta}}$ 
is sampled from the proposal density $q(\vecs{\theta} | \vecs{\theta}^j)$ and it is accepted with 
probability\cite{kupinski2003ideal}:
\begin{equation}
p_\text{a}(\vec{\hat{\vecs{\theta}}} | \vecs{\theta}^j, \vec{g}) = \min \left[1, \frac{p(\vec{g} | \vec{b}(\hat{\vecs{\theta}}), H_0) p(\hat{\vecs{\theta}}) q(\vecs{\theta}^j|\hat{\vecs{\theta}}) }{p(\vec{g} | \vec{b}({\vecs{\theta}}^j), H_0) p({\vecs{\theta}}^j) q(\hat{\vecs{\theta}} | \vecs{\theta}^j)}\right].
\end{equation}
The vector $\vecs{\theta}^{j+1} \equiv \hat{\vecs{\theta}}$ if the candidate is accepted; otherwise $\vecs{\theta}^{j+1} \equiv \vecs{\theta}^{j}$.
If the proposal density is designed to be symmetric, i.e., $q(\hat{\vecs{\theta}} | \vecs{\theta}^j) = q( \vecs{\theta}^j|\hat{\vecs{\theta} })$, the sampling strategy described above becomes a Metropolis-Hastings approach and the factors corresponding to the proposal density are cancelled.

Park  \emph{et al.} extended the MCMC approach to signal-known-statistically (SKS) signal detection tasks \cite{ park2003ideal} where the signal $\vec{s}$ is random. If the signal can be described by a set of stochastic parameters $\vecs{\alpha}$, i.e., $\vec{s} = \vec{s}(\vecs{\alpha})$, the likelihood ratio $\Lambda (\vec{g})$ can be written as\cite{ park2003ideal} :
\begin{equation}
\Lambda (\vec{g})  =\int d\vecs{\alpha} \int d\vecs{\theta}\ \Lambda_{\text{BSKE}} (\vec{g}|\vec{b}(\vecs{\theta}), \vec{s}(\vecs{\alpha}) ) p(\vecs{\theta}|\vec{g}, H_0) p(\vecs{\alpha}),
\end{equation}
where $\Lambda_{\text{BSKE}}(\vec{g}|\vec{b}(\vecs{\theta}), \vec{s}(\vecs{\alpha})  )=  \frac{p(\vec{g}|\vec{b}(\vecs{\theta}),\vec{s}(\vecs{\alpha}),H_1)}{p(\vec{g}|\vec{b}(\vecs{\theta}),H_0)}$.
The likelihood ratio can be subsequently approximated as: 
\begin{equation}
\hat{\Lambda}(\vec{g}) = \frac{1}{J}\sum_{j=1}^{J} \Lambda_{\text{BSKE}} (\vec{g}|\vec{b}(\vecs{\theta}^{(j)}), \vec{s}(\vecs{\alpha}^{(j)})).
\end{equation}
Here, $(\vecs{\theta}^j, \vecs{\alpha}^j)$ are sampled from the distribution $p(\vecs{\theta}| \vec{g}, H_0)p(\vecs{\alpha})$. The Markov chain can be constructed with acceptance probability:
\begin{equation} \label{eq:acp_sks}
p_\text{a}( \vec{\hat{\vecs{\theta}}},   \vec{\hat{\vecs{\alpha}}}| \vecs{\theta}^j, \vecs{\alpha}^j, \vec{g}) =
\min \left[1, \frac{p(\vec{g} | \vec{b}(\hat{\vecs{\theta}}), H_0) p(\hat{\vecs{\theta}}) p(\hat{\vecs{\alpha}})  q(\vecs{\theta}^j | \hat{\vecs{\theta}})q(\vecs{\alpha}^j | \hat{\vecs{\alpha}}) }{p(\vec{g} | \vec{b}({\vecs{\theta}}^j), H_0) p({\vecs{\theta}}^j)p(\vecs{\alpha}^j) q(\hat{\vecs{\theta}} | \vecs{\theta}^j)q(\hat{\vecs{\alpha}} | \vecs{\alpha}^j) }\right].
\end{equation}
Again, if the proposal densities are designed to be symmetric, the factors corresponding to the proposal density in Eq. \ref{eq:acp_sks} are canceled.

However, 
implementation of these proposed MCMC methods can be difficult because 
it is required to address practical issues such as the design of proposal density for the considered object model.
In addition, it remains unclear how to apply these methods under situations where the background cannot be described by 
the current well-established SOMs.

\subsection{Generative Adversarial Networks}
Generative Adversarial Networks (GANs) hold great promise to establish SOMs from training data\cite{goodfellow2014generative}.
A GAN trains a generator by competing against a discriminator through an adversarial process\cite{goodfellow2014generative}. 
When a GAN is trained on an ensemble of background images, 
the generator maps a random latent vector $\vec{z} \in \mathbb{R}^k$ to a synthetic background image $\hat{\vec{b}} = G(\vec{z}; \Theta_G)$.
Here, $G(\cdot\ ; \Theta_G): \mathbb{R}^k \rightarrow  \mathbb{R}^M$ is a mapping function represented by a deep neural network with a weight vector $\Theta_G$, and the latent vector $\vec{z}$ is sampled from a simple known distribution such as normal distribution or uniform distribution. 
The discriminator is represented by another deep neural network with a weight vector $\Theta_D$ and mapping function $D(\cdot\ ; \Theta_D): \mathbb{R}^M \rightarrow  \mathbb{R}$ that maps an image to a real-valued score to be used to distinguish between real and synthetic images. 
A GAN is trained by playing a two-player minimax game between the generator and the discriminator:
\begin{equation} \label{eq:GAN}
\min_{\Theta_G} \max_{\Theta_D} {E_{\vec{b} \sim p_b}} [l\left(D(\vec{b}; \Theta_D)\right)] + {E_{\vec{z}\sim p_z}} [l(1- D\left(G(\vec{z}; \Theta_G); \Theta_D\right) )],
\end{equation}
 where $l(\cdot)$ is an objective function, which is dependent on specific training strategies.
When $D(\cdot\ ; \Theta_D)$ and  $G(\cdot\ ; \Theta_G)$ possess sufficient capacity,  $p_{\hat{{b}}} = p_{{b}}$ when the global optimum of the minimax game is achieved\cite{goodfellow2014generative}.
Here, $p_{{b}}$ denotes the distribution of the real background images $\vec{b}$, and $p_{\hat{{b}}}$ denotes the distribution of the synthetic background images $\hat{\vec{b}}$.
The generator can subsequently represent a SOM that describes the variability within the ensemble of background objects.

\section{Markov-Chain Monte Carlo using Generative Adversarial Networks}
After a GAN has been trained on a set of background images, the distribution that describes the actual background images $p_{{{b}}}$ can be approximated by the 
distribution of GAN-produced background images $p_{{\hat{b}}}$: $p_{\hat{{b}}} \approx p_{{b}}$. 
The IO test statistic for SKE signal detection tasks can subsequently be approximated as:
\begin{equation}
\Lambda (\vec{g})  = \frac{\int d\hat{\vec{b}}\ p_{\hat{{b}}}(\hat{\vec{b}}) p(\vec{g}|\hat{\vec{b}}, H_1)}{\int d\hat{\vec{b}}\ p_{\hat{{b}}}(\hat{\vec{b}}) p(\vec{g}|\hat{\vec{b}}, H_0)} \equiv \int d \hat{\vec{b}}\ \Lambda_{\text{BKE}} (\vec{g}|\hat{\vec{b}}) p(\vec{\hat{{b}}}|\vec{g}, H_0),
\end{equation}
where $\Lambda_{\text{BKE}} (\vec{g}|\hat{\vec{b}}) = \frac{p(\vec{g}|\hat{\vec{b}},H_1)}{p(\vec{g}|\hat{\vec{b}},H_0)}$ and $p(\hat{\vec{b}}|\vec{g}, H_0) = {p(\vec{g}|\hat{\vec{b}}, H_0) p_{\hat{{b}}}(\hat{\vec{b}})} /{\int d\hat{\vec{b}}' \ p(\vec{g}|\hat{\vec{b}}', H_0) p_{\hat{{b}}}(\hat{\vec{b}}')}$.
Because $p(\hat{\vec{b}}|\vec{g},H_0) = \int d\vec{z}\ \delta(\hat{\vec{b}} - G(\vec{z};\Theta_G)) p(\vec{z}|\vec{g},H_0)$, where $\delta(\cdot)$ is a Dirac delta function and  $p(\vec{z}|\vec{g}, H_0) = \frac{p(\vec{g}|G(\vec{z};\Theta_G), H_0) p_z(\vec{z})}{\int d\vec{z'} p(\vec{g}|G(\vec{z'};\Theta_G), H_0) p_z(\vec{z'})}$, 
the likelihood ratio can be rewritten as:
\begin{equation}
\Lambda (\vec{g}) =\int d \hat{\vec{b}}\int d\vec{z}\ \Lambda_{\text{BKE}} (\vec{g}|\hat{\vec{b}}) \delta (\hat{\vec{b}} - G(\vec{z}; \Theta_G))p(\vec{z}|\vec{g}, H_0) = \int d\vec{z}\ \Lambda_{\text{BKE}} (\vec{g}|G(\vec{z}; \Theta_G)) p(\vec{z}|\vec{g}, H_0),
\end{equation}
where $\Lambda_{\text{BKE}} (\vec{g}|G(\vec{z}; \Theta_G))$ is evaluated on the synthetic background image generated by the GAN. 
The likelihood ratio subsequently can be approximated as:
\vspace{-0.1cm}
\begin{equation}
\hat{\Lambda}(\vec{g}) = \frac{1}{J}\sum_{j=1}^{J} \Lambda_{\text{BKE}} (\vec{g}|G(\vec{z}^j; \Theta_G)),
\end{equation}
where $\vec{z}^j$ is sampled from the posterior distribution $p(\vec{z}|\vec{g}, H_0)$. 
To produce $\vec{z}^j$, a Markov chain with an initial latent vector and a proposal density $q(\vec{z}|\vec{z}^j)$ can be constructed. Given $\vec{z}^j$, a candidate latent vector $\hat{\vec{z}}$ is drawn from the proposal density
and is accepted to the Markov chain with the acceptance probability: 
\begin{equation}
p_\text{a}(\vec{\hat{z}} | \vec{z}^j, \vec{g}) = \min \left[1, \frac{p\big(\vec{g} | G(\hat{\vec{z}};\Theta_G), H_0\big) p_z(\hat{\vec{z}}) q(\vec{z}^j|\hat{\vec{z}}) }{p\big(\vec{g} | G({\vec{z}}^j;\Theta_G), H_0\big) p_z({\vec{z}}^j) q(\hat{\vec{z}} | \vec{z}^j)}\right].
\end{equation}
Here, the probability density function $p_{z}(\cdot)$ has a simple analytical form because the latent vector $\vec{z}$ is sampled from simple distributions such
as the normal distribution or uniform distribution.
When a random walk Metropolis-Hastings (RWMH) algorithm\cite{pereyra2015survey} is employed, the proposal density $q(\hat{\vec{z}} | \vec{z}^j)$ can simply be chosen as a Gaussian density: $q(\hat{\vec{z}} | \vec{z}^j) \propto \exp{[-\frac{1}{2} (\hat{\vec{z}} - \vec{z}^j)^TK^{-1} (\hat{\vec{z}} - \vec{z}^j)]}$.
Additionally, because the gradient of the function represented by the generator $G({\vec{z}};\Theta_G)$
with respect to the latent vector $\vec{z}$ can be readily computed, more advanced MH algorithms including Metropolis adjusted Langevin algorithms (MALA)\cite{pereyra2015survey} and Hamiltonian
Monte Carlo (HMC)\cite{pereyra2015survey} that employ gradient information can be employed.

Similarly, the likelihood ratio for SKS signal detection tasks can be approximated as:
\begin{equation}
\hat{\Lambda}(\vec{g}) = \frac{1}{J}\sum_{j=1}^{J} \Lambda_{\text{BSKE}} (\vec{g}|G(\vec{z}^{(j)};\Theta_G), \vec{s}(\vecs{\alpha}^{(j)})),
\end{equation}
where $(\vec{z}^j, \vecs{\alpha}^j)$ are sampled from the distribution $p(\vec{z}| \vec{g}, H_0)p(\vecs{\alpha})$. 
A Markov chain for producing ($\vec{z}^j$, $\vecs{\alpha}^j$) can be constructed in a similar way as described above.

\section{Numerical studies}
The imaging system considered in this preliminary study was a parallel-hole collimator system whose point response function was described by a Gaussian kernel: $h_m(\vec{r}) = A\exp{\big[-\frac{(\vec{r} -\vec{r}_m )^T(\vec{r} -\vec{r}_m )}{2w^2}}\big]$. Here, the width $w = 0.5$ and the amplitude $A = \frac{h}{2\pi w^2}$ with the height $h = 40$. This imaging system recorded $64\times 64$ measured images.
A lumpy object model \cite{kupinski2001ideal} was considered with the mean number of lumps equal to 5. 
Each lump was represented by a Gaussian function with the width of 7 and amplitude of~1. 
Gaussian noise was considered with the standard deviation of 20 for the SKE detection task and 10 for the SKS detection task.
A deterministic signal described by a Gaussian function with the amplitude of 0.2 and the width of 3 was considered for the SKE signal detection task. 
For the SKS signal detection task, a uniformly distributed elliptical Gaussian signal with a random location and random shape was considered. 
We employed the method of Progressive Growing of GANs (ProGAN)\cite{karras2017progressive} and trained a ProGAN on 100,000 lumpy object images.
The ProGAN was trained in Tensorflow\cite{abadi2016tensorflow} by use of the Adam optimizer \cite{kingma2014adam}, which is a stochastic gradient algorithm.

\section{Results}
\begin{figure}[H]
\centering
{\includegraphics[width=0.8\linewidth]{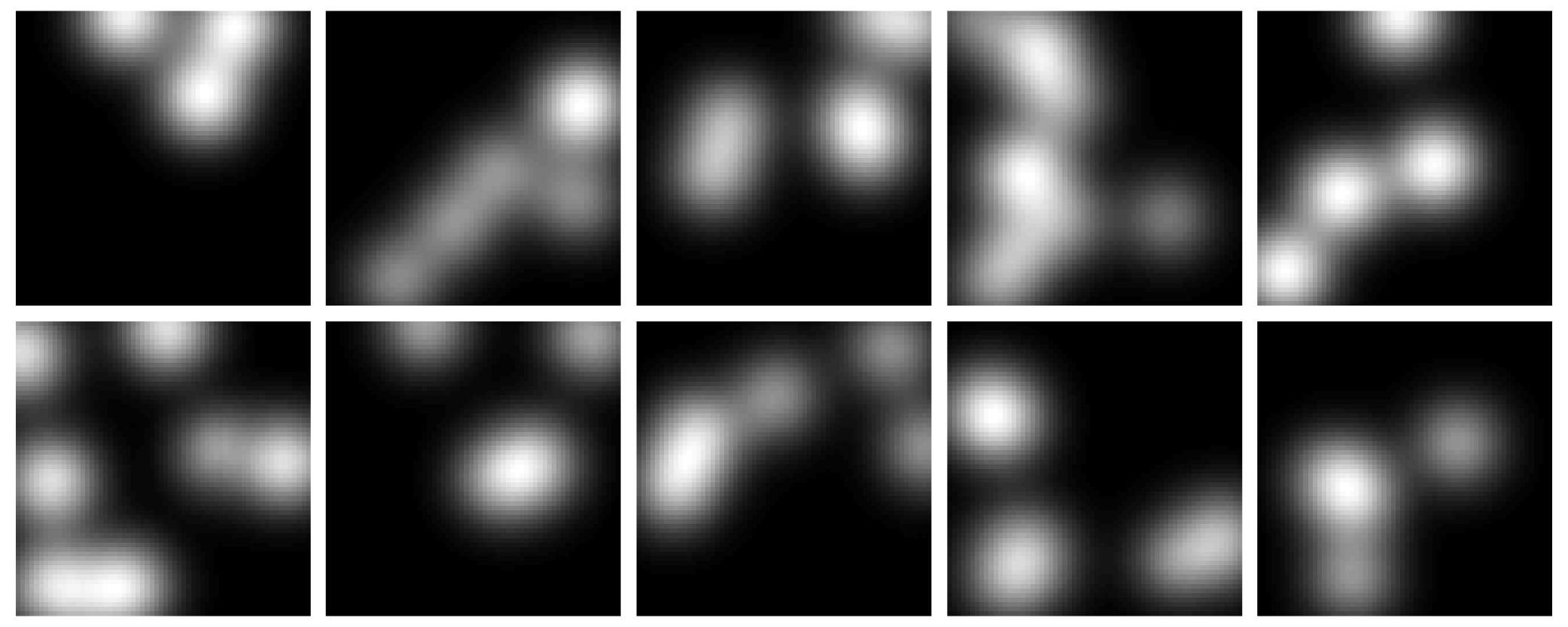}}
\vspace{.4cm}
\caption{ Top row: Samples of real background images; Bottom row: Samples of synthetic background images produced by a GAN.}
\label{fig:reals}
\end{figure}

Samples of real background images and synthetic background images produced by the GAN
are shown in Fig.~\ref{fig:reals}.

\begin{figure}[H]
\centering
 \begin{subfigure}{0.4\textwidth}
 \hspace{0cm} \subcaptionbox{}{\includegraphics[width=1.0\linewidth]{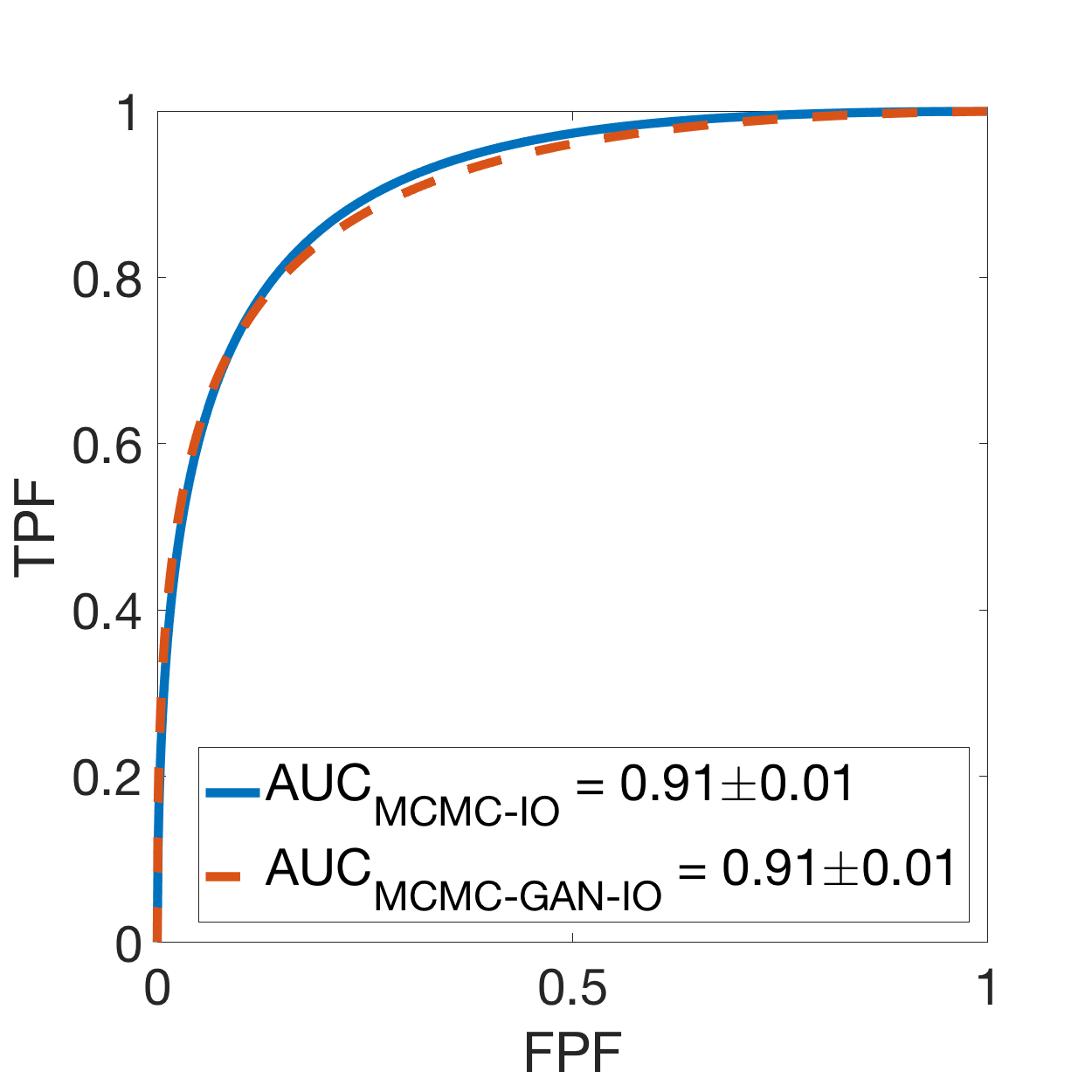}}
 \end{subfigure}
  \begin{subfigure}{0.4\textwidth}
 \hspace{0cm}\subcaptionbox{}{\includegraphics[width=1.0\linewidth]{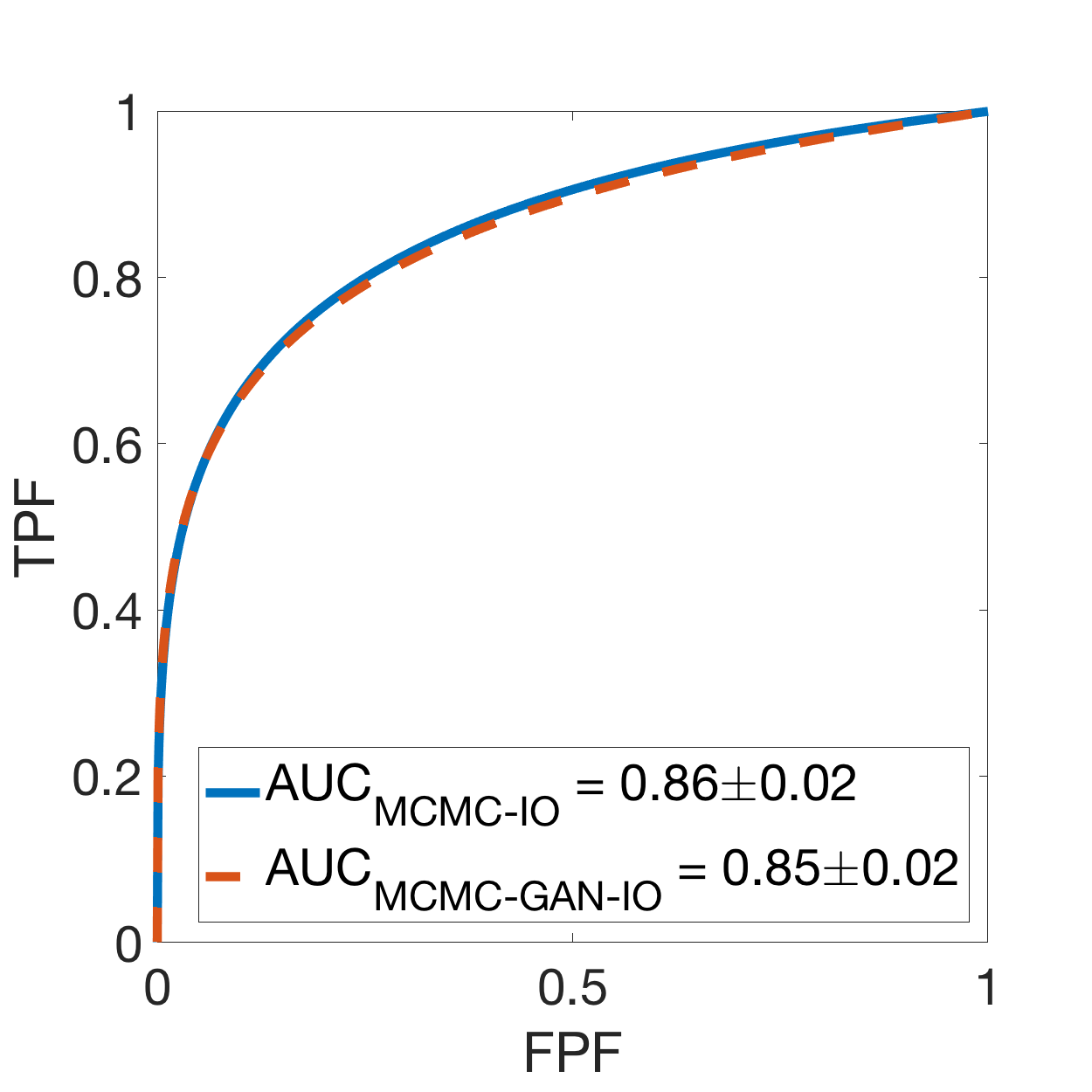}}
 \end{subfigure}
\caption{ (a) ROC curves and AUC values for the SKE detection task (b) ROC curves and AUC values for the SKS detection task.}
\label{fig:ROC}
\end{figure}

The IO performance was estimated by use of 200 pairs of signal-absent and signal-present images.
For each image, to generate a Markov chain, 100,000 iterations were run on the GAN-represented SOM by use of a RWMH algorithm with a simple Gaussian proposal density and the first 1000 burn-in iterations were ignored.
The IO performance estimated by use of the conventional MCMC algorithm that was applied to the original lumpy model was provided to validate the proposed method.
The ROC curves and AUC values corresponding to the MCMC-GAN method and the conventional MCMC method for the SKE detection task are shown in Fig. \ref{fig:ROC} (a),  and those for the SKS
detection task are shown in Fig. \ref{fig:ROC} (b).
The ROC curves were fit by use of the Metz-ROC software\cite{metz1998rockit} and the ``proper" binormal model\cite{pesce2007reliable}. 
In this proof-of-concept study where a simple lumpy object model was considered, the IO performance computed by our proposed MCMC-GAN method closely approximated that computed by the conventional MCMC method.
Because implementation of GANs is general and not limited to specific object models, our proposed method can be implemented with more sophisticated object models that can be learned by use of GANs where the conventional MCMC methods may not be available to use.

\section{Conclusion}
This work provides a novel methodology to approximate the IO performance by applying MCMC techniques with SOMs learned by use of GANs,
thereby extending the domain of applicability of MCMC methods. 
In this preliminary study, a lumpy background model was considered and
the IO performance computed by the proposed MCMC-GAN method 
closely approximated that computed by the conventional MCMC method for both the considered SKE and SKS signal detection tasks.
The proposed MCMC-GAN method can be potentially applied to more sophisticated object models learned by use of GANs.
This will enable computation of IO performance for optimizing imaging systems.
\section*{ACKNOWLEDGMENT}       
This research was supported in part by NIH awards EB020604, EB023045, NS102213, EB028652, and NSF award DMS1614305.

\bibliography{MCMCGAN.bib}
\bibliographystyle{spiebib} 

\end{document}